\documentclass[aps,prl,twocolumn,superscriptaddress]{revtex4-1}

\usepackage{graphicx}
\usepackage{amsmath,amsfonts,amssymb}
\usepackage[colorlinks=true,linkcolor=blue,citecolor=blue]{hyperref}

\usepackage{siunitx}
\usepackage{layouts}
\usepackage{todonotes}

\begin{document}

\title{Large spin relaxation anisotropy and valley-Zeeman spin-orbit coupling in WSe2/Gr/hBN heterostructures}


\author{Simon Zihlmann}
\email{Simon.Zihlmann@unibas.ch}
\affiliation{Department of Physics, University of Basel, Klingelbergstrasse 82, CH-4056 Basel, Switzerland}

\author{Aron W. Cummings}
\affiliation{Catalan Institute of Nanoscience and Nanotechnology (ICN2), CSIC and BIST, Campus UAB, Bellaterra, 08193 Barcelona, Spain\\}

\author{Jose H. Garcia}
\affiliation{Catalan Institute of Nanoscience and Nanotechnology (ICN2), CSIC and BIST, Campus UAB, Bellaterra, 08193 Barcelona, Spain\\}

\author{M\'at\'e Kedves}
\affiliation{Department of Physics, Budapest University of Technology and Economics and Nanoelectronics 'Momentum' Research Group of the Hungarian Academy of Sciences, Budafoki ut 8, 1111 Budapest, Hungary}

\author{Kenji Watanabe}
\affiliation{National Institute for Material Science, 1-1 Namiki, Tsukuba, 305-0044, Japan\\}

\author{Takashi Taniguchi}
\affiliation{National Institute for Material Science, 1-1 Namiki, Tsukuba, 305-0044, Japan\\}

\author{Christian Sch\"onenberger}
\affiliation{Department of Physics, University of Basel, Klingelbergstrasse 82, CH-4056 Basel, Switzerland}

\author{P\'eter Makk}
\email{peter.makk@mail.bme.hu}
\affiliation{Department of Physics, University of Basel, Klingelbergstrasse 82, CH-4056 Basel, Switzerland}
\affiliation{Department of Physics, Budapest University of Technology and Economics and Nanoelectronics 'Momentum' Research Group of the Hungarian Academy of Sciences, Budafoki ut 8, 1111 Budapest, Hungary}

\date{\today}
\begin{abstract}
Large spin-orbital proximity effects have been predicted in graphene interfaced with a transition metal dichalcogenide layer. Whereas clear evidence for an enhanced spin-orbit coupling has been found at large carrier densities, the type of spin-orbit coupling and its relaxation mechanism remained unknown. We show for the first time an increased spin-orbit coupling close to the charge neutrality point in graphene, where topological states are expected to appear. Single layer graphene encapsulated between the transition metal dichalcogenide WSe$_2$ and hBN is found to exhibit exceptional quality with mobilities as high as \SI{100000}{\square\centi\metre\per\volt\per\second}. At the same time clear weak anti-localization indicates strong spin-orbit coupling and a large spin relaxation anisotropy due to the presence of a dominating symmetric spin-orbit coupling is found. Doping dependent measurements show that the spin relaxation of the in-plane spins is largely dominated by a valley-Zeeman spin-orbit coupling and that the intrinsic spin-orbit coupling plays a minor role in spin relaxation. The strong spin-valley coupling opens new possibilities in exploring spin and valley degree of freedom in graphene with the realization of new concepts in spin manipulation.
\end{abstract}

\maketitle

\section{Motivation/Introduction}
\label{sec:Introduction}

In recent years, van der Waals heterostructures (vdW) have gained a huge interest due to their possibility of implementing new functionalities in devices by assembling 2D building blocks on demand \cite{2013_Geim}. It has been shown that the unique band structure of graphene can be engineered and enriched with new properties by placing it in proximity to other materials, including the formation of minibands \cite{2013_Ponomarenko, 2013_Dean, 2013_Hunt, 2016_lee}, magnetic ordering \cite{2015_Wang_b, 2017_Leutenantsmeyer}, and superconductivity \cite{2015_Efetov, 2017_Bretheau}. Special interest has been paid to the enhancement of spin-orbit coupling (SOC) in graphene since a topological state, a quantum spin Hall phase, was theoretically shown to emerge \cite{2005_Kane}. First principles calculations predicted an intrinsic SOC strength of \SI{12}{\micro\eV} \cite{2010_Konschuh}, which is currently not observable even in the cleanest devices. Therefore, several routes were proposed and explored to enhance the SOC in graphene while preserving its high electronic quality \cite{2009_CastroNeto, 2014_Han, 2015_Gmitra}. One of the most promising approaches is the combination of a transition metal dichalcogenide (TMDC) layer with graphene in a vdW-hetereostructure. TMDCs have very large SOC on the \SI{100}{\milli\eV}--scale in the valence band and large SOC on the order of \SI{10}{\milli\eV} in the conduction band \cite{2014_Han}.

The realization of topological states is not the only motivation to enhance the SOC in graphene. It has been shown that graphene is an ideal material for spin transport \cite{2014_Han}. Spin relaxation times on the order of nanoseconds \cite{2014_Droegeler, 2016_Singh} and relaxation lengths of \SI{24}{\micro\metre} \cite{2015_Ingla-Aynes} have been observed. However, the presence of only weak SOC in pristine graphene limits the tunability of possible spintronics devices made from graphene. The presence of strong SOC would enable fast and efficient spin manipulation by electric fields for possible spintronics applications, such as spin-filters \cite{2017_Cummings} or spin-orbit valves \cite{2017_Gmitra, 2017_Khoo}. In addition, enhanced SOC leads to large spin-Hall angles \cite{2017_Garcia} that could be used as a source of spin currents or as a detector of spin currents in graphene-based spintronic devices. 

It was proposed that graphene in contact to a single layer of a TMDC can inherit a substantial SOC from the underlying substrate \cite{2015_Gmitra, 2016_Gmitra}. The experimental detection of clear weak anti-localization (WAL) \cite{2015_Wang_a, 2016_Wang, 2016_Yang, 2017_Voelkl, 2017_Yang, 2017_Wakamura} as well as the observation of a beating of Shubnikov de-Haas (SdH) oscillations \cite{2016_Wang} leave no doubt that the SOC is greatly enhanced in graphene/TMDC heterostructures. First principles calculations of graphene on WSe$_2$ \cite{2016_Gmitra} predicted large spin-orbit coupling strength and the formation of inverted bands hosting special edge states. At low energy, the band structure can be described in a simple tight-binding model of graphene containing the orbital terms and all the symmetry allowed SOC terms $H = H_0 + H_\Delta + H_I + H_{VZ} + H_R$ \cite{2016_Gmitra, 2017_Kochan}:
\begin{equation}
	\label{eq:Hamiltonian}
	\begin{aligned}
	&H_0 = \hbar v_F\left(\kappa k_x\hat{\sigma_x} + k_y\hat{\sigma_y}\right)\cdot \hat{s_0} \\
	&H_\Delta = \Delta\hat{\sigma}_z\cdot\hat{s}_0 \\
	&H_I = \lambda_I\kappa\hat{\sigma}_z\cdot\hat{s}_z\\
	&H_{VZ} = \lambda_{VZ}\kappa\hat{\sigma}_0\cdot\hat{s}_z \\
	&H_R = \lambda_R\left(\kappa\hat{\sigma}_x\cdot\hat{s}_y - \hat{\sigma}_y\cdot\hat{s}_x\right). \\
	\end{aligned}
\end{equation}
Here, $\hat{\sigma_i}$ are the Pauli matrices acting on the pseudospin, $\hat{s_i}$ are the Pauli matrices acting on the real spin and $\kappa$ is either $\pm1$ and denotes the valley degree of freedom. $k_x$ and $k_y$ represent the k-vector in the graphene plane, $\hbar$ is the reduced Planck constant, $v_F$ is the Fermi velocity and $\lambda_i, \Delta$ are constants. The first term $H_0$ is the usual graphene Hamiltonian that describes the linear band structure at low energies. $H_\Delta$ represents an orbital gap that arises from a staggered sublattice potential. $H_I$ is the intrinsic SOC term that opens a topological gap of $2\lambda_I$ \cite{2005_Kane}. $H_{VZ}$ is a valley-Zeeman SOC that couples valley to spin and results from different intrinsic SOC on the two sublattices. This term leads to a Zeeman splitting of $2\lambda_{VZ}$ that has opposite sign in the K and K' valleys and leads to an out of plane spin polarization with opposite polarization in each valley. $H_R$ is a Rashba SOC arising from the structure inversion asymmetry. This term leads to a spin splitting of the bands with a spin expectation value that lies in the plane and is coupled to the momentum via the pseudospin. At higher energies k-dependent terms, called pseudospin inversion asymmetric (PIA) SOC come into play, which can be neglected at lower doping \cite{2017_Kochan}.

Previous studies have estimated the SOC strength from theoretical calculations \cite{2015_Wang_a} or extracted only the Rashba SOC at intermediate \cite{2017_Yang} or at very high doping \cite{2016_Yang} or gave only  a total SOC strength \cite{2017_Voelkl}. Further studies have extracted a combination of Rashba and valley-Zeeman SOC strength form SdH-oscillation beating measurements \cite{2016_Wang}. Additionally, a very recent study uses the clean limit (precession time) to estimate the SOC strength from diffusive WAL measurements \cite{2017_Wakamura}.

Here, we give for the first time a clear and comprehensive study of SOC at the charge neutrality point (CNP) for WSe$_2$/Gr/hBN heterostructures. The influence of strong SOC is expected to have the largest impact on the bandstructure close to the CNP. The strength of all possible SOC terms is discussed and we find that the relaxation times are dominated by the valley-Zeeman SOC. The valley-Zeeman SOC leads to a much faster relaxation of in-plane spins than out-of plane spins. This asymmetry is unique for systems with strong valley-Zeeman SOC and is not present in traditional 2D Rashba systems where the anisotropy is 1/2 \cite{2017_Cummings}. Our study is in contrast to previous WAL measurements \cite{2016_Yang, 2017_Yang}, but is in good agreement with recent spin-valve measurements reporting a large spin relaxation anisotropy \cite{2017_Ghiasi, 2017_Benitez}.

\section{Methods}
\label{sec:Methods}

WSe$_2$/Gr/hBN vdW-heterostructures were assembled using a dry pick-up method \cite{2014_Zomer} and Cr/Au 1D-edge contacts were fabricated \cite{2013_Wang_a}. Obviously a clean interface between high quality WSe$_2$ and graphene is of utmost importance. A short discussion on the influence of the WSe$_2$ quality is given in the Supplemental Material. After shaping the vdW-heterostructure into a Hall-bar geometry by a reactive ion etching plasma employing SF$_6$ as the main reactive gas, Ti/Au top gates were fabricated with an MgO dielectric layer to prevent it from contacting the exposed graphene at the edge of the vdW-heterostructure. A heavily-doped silicon substrate with \SI{300}{\nano\metre} SiO$_2$ was used as a global back gate. An optical image of a typical device and a cross section is shown in Fig.~\ref{fig:device}~\textbf{(a)}. In total, three different samples with a total of four devices were fabricated. Device A, B and C are presented in the main text and device D is discussed in the Supplemental Material. Standard low frequency lock-in techniques were used to measure two- and four-terminal conductance and resistance. Weak anti-localization was measured at temperatures of \SI{50}{\milli\kelvin} to \SI{1.8}{\kelvin} whereas a classical background was measured at sufficiently large temperatures of \SIrange{30}{50}{\kelvin}.

\section{Results}
\label{sec:Results}

\subsection{Device Characterization}
\begin{figure}[htbp]
	\includegraphics[scale=1]{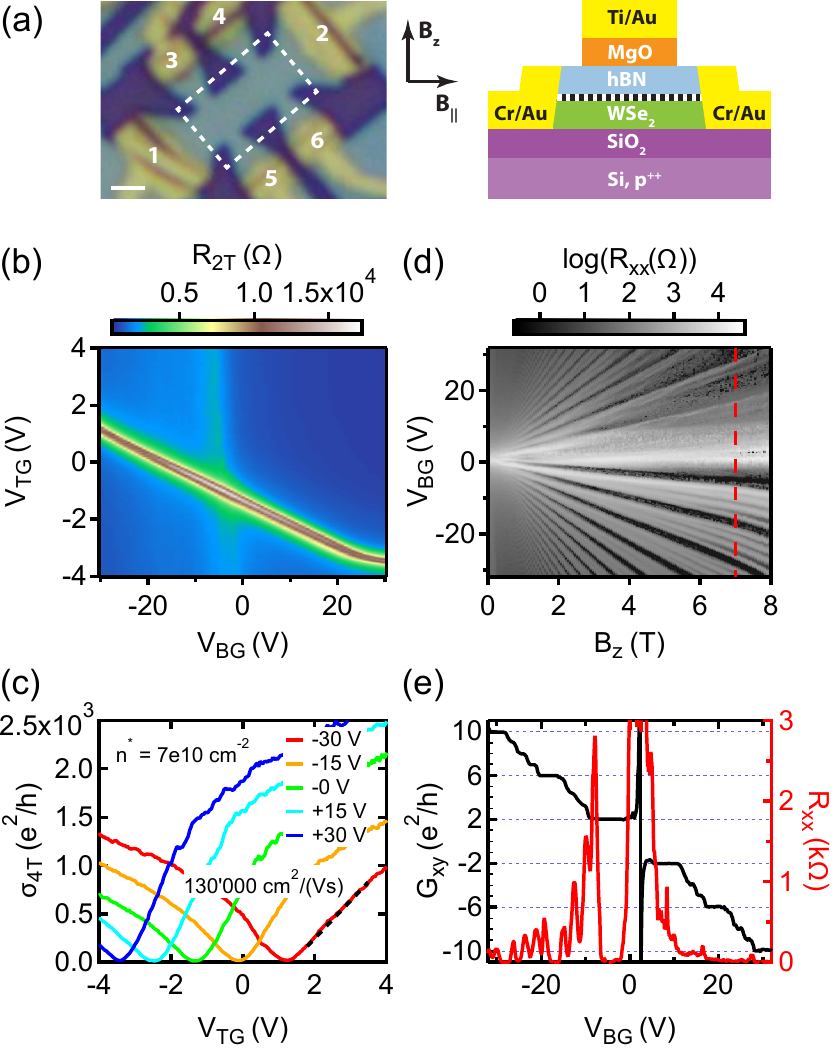}
	\caption{\label{fig:device}\textbf{Device layout and basic characterization of WSe$_2$/Gr/hBN vdW-heterostructures.} \textbf{(a)} shows an optical image of a device A before the fabrication of the top gate, whose outline is indicated by the white dashed rectangle. On the right, a schematic cross section is shown and the directions of the magnetic fields are indicated. The scale bar is \SI{1}{\micro\metre}. The data shown in \textbf{(b)} to \textbf{(e)} are from device B. The two terminal resistance measured from lead 1 to 2 is shown as a function of top and back gate voltage. A pronounced resistance maximum tunable by both gates indicates the charge neutrality point (CNP) of the bulk device, whereas a fainter line only changing with V$_\mathrm{BG}$ indicates the CNP from the device area close to the contacts that are not covered by the top gate. Cuts in V$_\mathrm{TG}$ at different V$_\mathrm{BG}$ of the conductivity measured in a four-terminal configuration are shown in \textbf{(c)}, which are also used to extract field effect mobility (linear fit indicated by black dashed line) and residual doping as indicated. The fan plot of longitudinal resistance R$_{\mathrm{xx}}$ versus V$_\mathrm{BG}$ and B$_\mathrm{z}$ at V$_\mathrm{TG}$ = \SI{-1.42}{\volt} is shown in \textbf{(d)} and a cut at B$_\mathrm{z}$ = \SI{7}{\tesla} in \textbf{(e)}. Clear plateaus are observed at filling factors $\nu=\pm2, \pm3, \pm4, \dots$ and higher, indicating full lifting of the fourfold degeneracy of graphene for magnetic fields $>$~\SI{6}{\tesla}.}
\end{figure}

The two-terminal resistance measured from contact 1 to 2 as a function of applied top and bottom gate is shown in Fig. \ref{fig:device} \textbf{(b)}. A pronounced resistance maximum, tunable by both gates, indicates the CNP of the bulk of the device whereas a fainter line only changing with V$_\mathrm{BG}$ indicates the CNP from the device areas close to the contacts, which are not covered by the top gate. From the four-terminal conductivity, shown in Fig.~\ref{fig:device}~\textbf{(c)}, the field effect mobility $\mu\simeq$~\SI{130000}{\square\centi\metre\per\volt\per\second} and the residual doping $n^*$~=~\SI{7e10}{\square\per\centi\metre} were extracted. The mobility was extracted from a linear fit of the conductivity as a function of density at negative V$_\mathrm{BG}$. At positive V$_\mathrm{BG}$ the mobility is higher as one can easily see from Fig.~\ref{fig:device}~\textbf{(c)}. At $\mathrm{V_{BG}}~\geq$~\SI{25}{\volt}, the lever arm of the back gate is greatly reduced since the WSe$_2$ layers gets populated with charge carriers, i.g. the Fermi level is shifted into some trap states in the WSe$_2$. Although the WSe$_2$ is poorly conducting (low mobility) it can screen potential fluctuations due to disorder and this can lead to a larger mobility in the graphene layer, as similarly observed in graphene on MoS$_2$ \cite{2017_Banszerus}.

Fig.~\ref{fig:device}~\textbf{(d)} shows the longitudinal resistance as a function of magnetic field and gate voltage with lines originating from the integer quantum Hall effect. At low fields, the normal single layer spectrum is obtained with plateaus at filling factors $\nu = \pm2, \pm6, \pm10, \pm14, \dots$, whereas at larger magnetic fields full degeneracy lifting is observed with plateaus at filling factors $\nu = \pm2, \pm3, \pm4, \pm5, \pm6, \dots$. The presence of symmetry broken states, that are due to electron-electron interactions \cite{2012_Young}, is indicative of a high device quality. In the absence of interaction driven symmetry breaking, the spin-splitting of the quantum Hall states could be used to investigate the SOC strength \cite{2017_Cysne}.

The high quality of the devices presented here poses sever limitations on the investigation of the SOC strength using WAL theory. Ballistic transport features (transverse magnetic focusing) are observed at densities larger than \SI{8e11}{\per\square\centi\metre}. Therefore, a true diffusive regime is only obtained close to the CNP, where the charge carriers are quasi-diffusive \cite{2011_DasSarma}.

\label{subsec:device}

\subsection{Magneto conductance}
\label{subsec:magneto_conductance}

In a diffusive conductor, the charge carrier trajectories can form closed loops after several scattering events. The presence of time-reversal symmetry leads to a constructive interference of the electronic wave function along these trajectories and therefore to an enhanced back scattering probability compared to the classical case. This phenomenon is known as weak localization (WL). Considering the spin degree of freedom of the electrons, this can change. If strong SOC is present the spin can precess between scattering events, leading to destructive interference and hence to an enhanced forward scattering probability compared to the classical case. This phenomenon is known as weak anti-localization \cite{1982_Bergmann}. The quantum correction to the magneto conductivity can therefore reveal the SOC strength.

\begin{figure}[htbp]
	\includegraphics[scale=1]{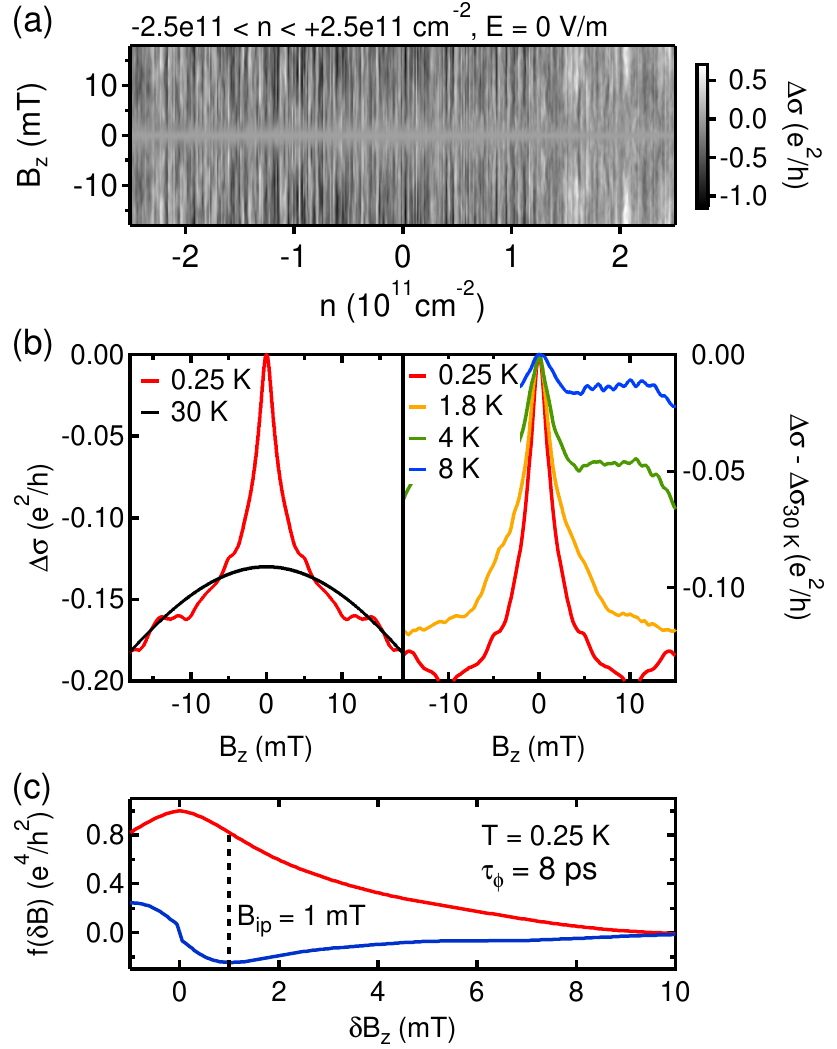}
	\caption{\label{fig:magneto_conductance}\textbf{Magneto conductivity of device A:} \textbf{(a)} Magneto conductivity versus B$_\mathrm{z}$ and n is shown at T = \SI{0.25}{\kelvin}. A clear feature is observed around B = \SI{0}{\milli\tesla} and large modulations due do UCF are observed in B$_\mathrm{z}$ and n. \textbf{(b)} shows the magneto conductivity averaged over all traces at different n. The WAL peak completely disappears at T = \SI{30}{\kelvin}, leaving the classical magneto conductivity as a background. The \SI{30}{\kelvin} trace is offset vertically for clarity. The quantum correction to the magneto conductivity is then obtained by subtracting the high temperature background from the magneto conductivity, see \textbf{(b)} on the right for different temperatures. With increasing temperature the phase coherence time shortens and therefore the WAL peak broadens and reduces in height. \textbf{(c)} shows the autocorrelation of the magneto conductivity in red and its derivative in blue (without scale). The minimum of the derivative indicates the inflection point (B$_\mathrm{{ip}}$) of the autocorrelation, which is a measure of $\tau_\phi$. }
\end{figure}

The two-terminal magneto conductivity $\Delta\sigma = \sigma\left(B\right) - \sigma\left(B=0\right)$ versus B$_\mathrm{z}$ and n at T = \SI{0.25}{\kelvin} and zero perpendicular electric field is shown in Fig.~\ref{fig:magneto_conductance}~\textbf{(a)}. A clear feature at B$_\mathrm{z}$~=~\SI{0}{\milli\tesla} is visible, as well as large modulations in B$_\mathrm{z}$ and n due to universal conductance fluctuations (UCFs). UCFs are not averaged out since the device size is on the order of the dephasing length $l_\phi$. Therefore, an ensemble average of the magneto conductivity over several densities is performed to reduce the amplitude of the UCFs \cite{2015_Wang_a}, and curves as in Fig.~\ref{fig:magneto_conductance}~\textbf{(b)} result. A clear WAL peak is observed at \SI{0.25}{\kelvin} whereas at \SI{30}{\kelvin} the quantum correction is fully suppressed due to a very short phase coherence time and only a classical background in magneto conductivity remains. This high temperature background is then subtracted from the low temperature measurements to extract the real quantum correction to the magneto conductivity \cite{2016_Wang}. In addition to WL/WAL measurements the phase coherence time can be extracted independently from the autocorrelation function of UCF in magnetic field \cite{1987_Lee}. UCF as a function of B$_\mathrm{z}$ was measured in a range where the WAL did not contribute to the magneto conductivity (e.g. \SIrange{20}{70}{\milli\tesla}) and an average over several densities was performed. The inflection point in the autocorrelation, determined by the minimum in its derivative, is a robust measure of $\tau_\phi$ \cite{2012_Lundeberg}, see Fig.~\ref{fig:magneto_conductance}~\textbf{(d)}.

\subsection{Fitting}
\label{subsec:fitting}
To extract the spin-orbit scattering times we use the theoretical formula derived by diagrammatic perturbation theory \cite{2012_McCann}. In the case of graphene, the quantum correction to the magneto conductivity $\Delta\sigma$ in the presence of strong SOC is given by: 
\begin{equation}
	\label{eq:WAL}
	\begin{split}
	\Delta\sigma (B) = -\frac{e^2}{2\pi h}\left[ F \left( \frac{\tau_B^{-1}}{\tau_\phi^{-1}} \right) - F \left( \frac{\tau_B^{-1}}{\tau_\phi^{-1} + 2\tau_{asy}^{-1}} \right) \right. \\
	\left. - 2F \left( \frac{\tau_B^{-1}}{\tau_\phi^{-1} + \tau_{asy}^{-1} + \tau_{sym}^{-1}} \right) \vphantom{\int_1^2} \right],
	\end{split}
\end{equation}
where $F(x) = \ln(x) + \Psi(1/2+1/x)$, with $\Psi(x)$ being the digamma function, $\tau_B^{-1} = 4eDB/\hbar$, where $D$ is the diffusion constant, $\tau_\phi$ is the phase coherence time, $\tau_{asy}$ is the spin-orbit scattering time due to SOC terms that are asymmetric upon z/-z inversion ($H_R$) and $\tau_{sym}$ is the spin-orbit scattering time due to SOC terms that are symmetric upon z/-z inversion ($H_I$, $H_{VZ}$) \cite{2012_McCann}. The total spin-orbit scattering time is given by the sum of the asymmetric and symmetric rate $\tau_{SO}^{-1} = \tau_{asy}^{-1} + \tau_{sym}^{-1}$. In general, Eq. \ref{eq:WAL} is only valid if the intervalley scattering rate $\tau_{iv}^{-1}$ is much larger than the dephasing rate $\tau_\phi^{-1}$ and the rates due to spin-orbit scattering $\tau_{asy}^{-1}$, $\tau_{sym}^{-1}$.

In the limit of very weak asymmetric but strong symmetric SOC ($\tau_{asy} \gg \tau_\phi \gg \tau_{sym}$), Eq. \ref{eq:WAL} describes reduced WL since the first two terms cancel and therefore a positive magneto conductivity results. Contrary to that, in the limit of very weak symmetric but strong asymmetric SOC ($\tau_{sym} \gg \tau_\phi \gg \tau_{asy}$) a clear WAL peak is obtained. If both time scales are shorter than $\tau_\phi$, the ratio $\tau_{asy}/\tau_{sym}$ will determine the quantum correction of the magneto conductivity. In the limit of total weak SOC ($\tau_{asy}, \tau_{sym} \gg \tau_\phi$) the normal WL in graphene is obtained \cite{2006_McCann}, as the first two terms cancel and other terms explicitly involving the inter- and intravalley scattering must be considered (see Supplemental Material).

Since the second and the third term can produce very similar dependencies on B$_\mathrm{z}$ it can be hard to properly distinguish between the influence of $\tau_{asy}$ and $\tau_{sym}$ on $\Delta\sigma(B)$, as also previously reported \cite{2016_Wang, 2017_Wakamura}. It is therefore important to measure and fit the magneto conductivity to sufficiently large fields in order to capture the influence of the second and third term, which only significantly contribute at larger fields (for strong SOC). However, there is an upper limit of the field scale (the so-called transport field $B_{tr}$) at which the theory of WAL breaks down. The size of the shortest closed loops that can be formed in a diffusive sample is on the order of $l_{mfp}^2$, where $l_{mfp}$ is the mean-free path of the charge carriers. Fields that are larger than $\Phi_0 / l_{mfp}^2$, where $\Phi_0 = h/e$ is the flux quantum, are not meaningful in the framework of diffusive transport.

\begin{figure}[htbp]
	\includegraphics[scale=1]{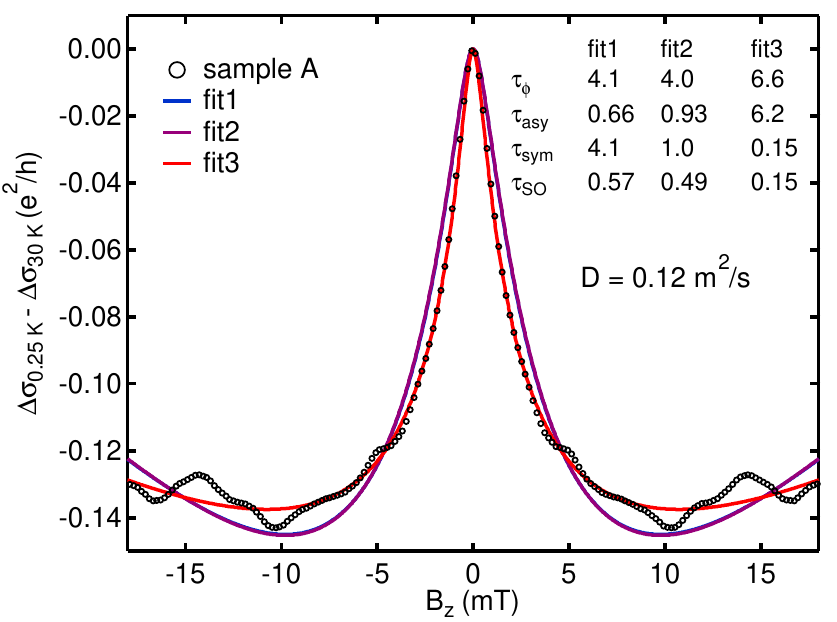}
	\caption{\label{fig:fitting}\textbf{Fitting of quantum correction to the magneto conductivity of device A} The quantum correction to the magneto conductivity is fit using Eq. \ref{eq:WAL}. The results for three different limits are shown and their parameters are indicated (in units of \si{\pico\second}). $\tau_\phi$ is estimated to be \SI{8}{\pico\second} from the autocorrelation of UCF in magnetic field, see Fig. \ref{fig:magneto_conductance} \textbf{(d)}.}
\end{figure}

In the most general case there are three different regimes in the presence of strong SOC in graphene: $\tau_{asy} \ll \tau_{sym}$, $\tau_{asy} \sim \tau_{sym}$ and $\tau_{asy} \gg \tau_{sym}$. Therefore, we fitted the magneto conductivity with initial fit parameters in these three limits. An example is shown in Fig. \ref{fig:fitting}, where the three different fits are shown as well as the extracted parameters. Obviously, the case $\tau_{asy} \gg \tau_{sym}$ (fit1) and  $\tau_{asy} \sim \tau_{sym}$ (fit2) are indistinguishable and fit the data worse than the case $\tau_{asy} \gg \tau_{sym}$ (fit3). In addition, $\tau_\phi$ extracted from the UCF matches best for fit3. Therefore, we can clearly state that the symmetric SOC is stronger than the asymmetric SOC. The flat background as well as the narrow width of the WAL peak can only be reproduced with the third case. A very similar behaviour was found in device C at the CNP. In device B (shown in the Supplemental Material), whose mobility is larger than the one from device A, we cannot clearly distinguish the three limits as the transport field is too low ($\approx$~\SI{12}{\milli\tesla}) and the flat background at larger field cannot be used to disentangle the different parameters from each other. However, this does not contradict $\tau_{asy} \gg \tau_{sym}$ and the overall strength of the SOC ($\tau_{SO}\simeq$~\SI{0.2}{\pico\second}) is in good agreement with device A shown here.

Obviously, the extracted time scales should be taken with care as many things can introduce uncertainties in the extracted time scales. First of all, we are looking at ensemble-averaged quantities and it is clear that this might influence the precision of the extraction of the time scales. In addition, the subtraction of a high temperature background can lead to higher uncertainty of the quantum correction. Lastly, the high mobility of the clean devices places severe limitations on the usable range of magnetic field. All these influences lead us to a conservative estimation of a \SI{50}{\percent} uncertainty for the extracted time scales. Nevertheless, the order of magnitude of the extracted time scales and trends are still robust.

The presence of a top and a back gate allows us to tune the carrier density and the transverse electric field independently. The spin-orbit scattering rates were found to be electric field independent at the CNP in the range of \SIrange{-0.05}{0.08}{\volt\per\nano\metre} within the precision of parameter extraction. Details are given in the Supplemental Material. Within the investigated electric field range $\tau_{asy}$ was found to be in the range of \SIrange{5}{10}{\pico\second}, always close to $\tau_\phi$. $\tau_{sym}$ on the other hand was found to be around \SIrange{0.1}{0.3}{\pico\second} while $\tau_p$ was around \SIrange{0.2}{0.3}{\pico\second}, see Supplemental Material for more details. The lack of electric field tunability of $\tau_{asy}$ and $\tau_{sym}$ in the investigated electric field range is not so surprising.  The Rashba coupling in this system is expected to change considerably for electric fields on the order of \SI{1}{\volt\per\nano\metre}, which are much larger than the applied fields here. However, such large electric fields are hard to achieve. In addition, $\tau_{sym}$, which results from $\lambda_I$ and $\lambda_{VZ}$ is not expected to change much with electric field as long as the Fermi energy is not shifted into the conduction or valence band of the WSe$_2$ \cite{2015_Gmitra}. These findings contradict another study \cite{2017_Voelkl}, which claims an electric field tunability of both SOC terms. However, there it is not discussed how accurately those parameters were extracted.

\subsection{Density dependence}
\begin{figure}[htbp]
	\includegraphics[scale=1]{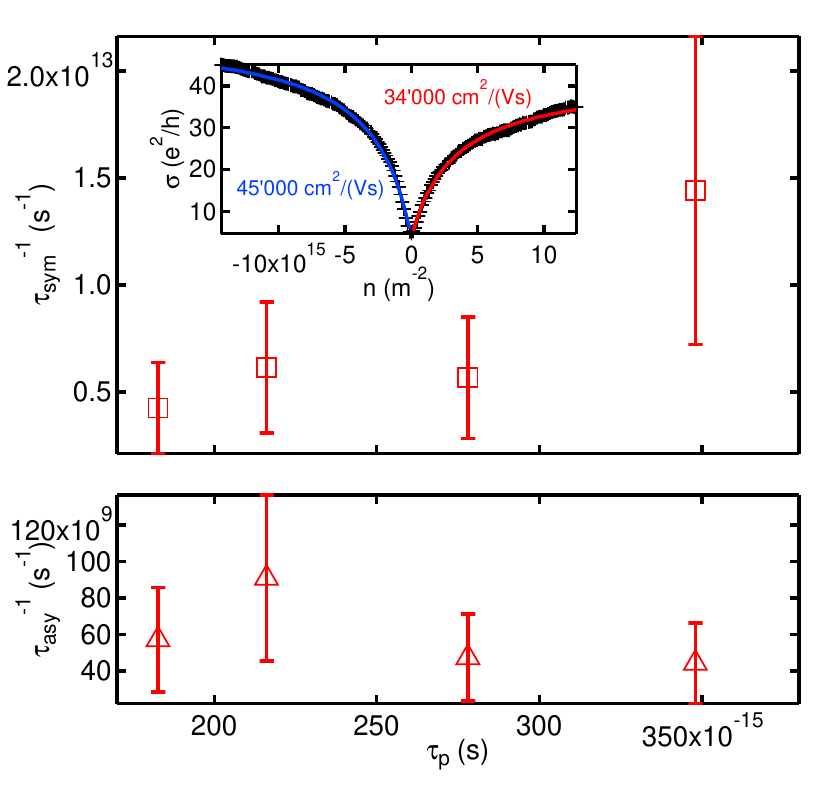}
	\caption{\label{fig:n_dep}\textbf{Density dependence of device C:} The dependence of the spin-orbit scattering rates $\tau_{sym}^{-1}$ and $\tau_{asy}^{-1}$ as a function of $\tau_p$ are shown for device C. The error bars on the spin-orbit scattering rates are given by a conservative estimate of \SI{50}{\percent}. The two terminal conductivity is shown in the inset and the extracted mobilities for the n and p side are indicated. }
\end{figure}

The momentum relaxation time $\tau_p$ can be tuned by changing the carrier density in graphene. Fig.~\ref{fig:n_dep} shows the dependence of $\tau_{asy}^{-1}$ and $\tau_{sym}^{-1}$ on $\tau_p$ in a third device C. The lower mobility in device C allowed for WAL measurements at higher charge carrier densities not accessible in devices A and B. At the CNP, $\tau_{asy}^{-1}$ and $\tau_{sym}^{-1}$ are found to be consistent across all three devices A, B and C. Here, $\tau_{sym}^{-1}$ increases with increasing $\tau_p$ whereas $\tau_{asy}^{-1}$ is roughly constant with increasing $\tau_p$. The dependence of the spin-orbit scattering times on the momentum scattering time can give useful insights into the dominating spin relaxation mechanisms, as will be discussed later. It is important to note that the extracted $\tau_{asy}$ is always very close to $\tau_\phi$. Therefore, the extracted $\tau_{asy}$ could be shorter than what the actual value would be since $\tau_\phi$ acts as a cutoff.

\label{subsec:n_dep}

\subsection{In-plane magnetic field dependence}
\label{subsec:B_ip_dep}
An in-plane magnetic field (B$_\parallel$) is expected to lift the influence of SOC on the quantum correction to the magneto conductivity at sufficiently large fields. This means that a crossover from WAL to WL for z/-z asymmetric and a crossover from reduced WL to full WL correction for z/-z symmetric spin-orbit coupling is expected at a field where the Zeeman energy is much larger than the SOC strength \cite{2012_McCann}. The experimental determination of this crossover field allows for an estimate of the SOC strength.

\begin{figure}[htbp]
	\includegraphics[scale=1]{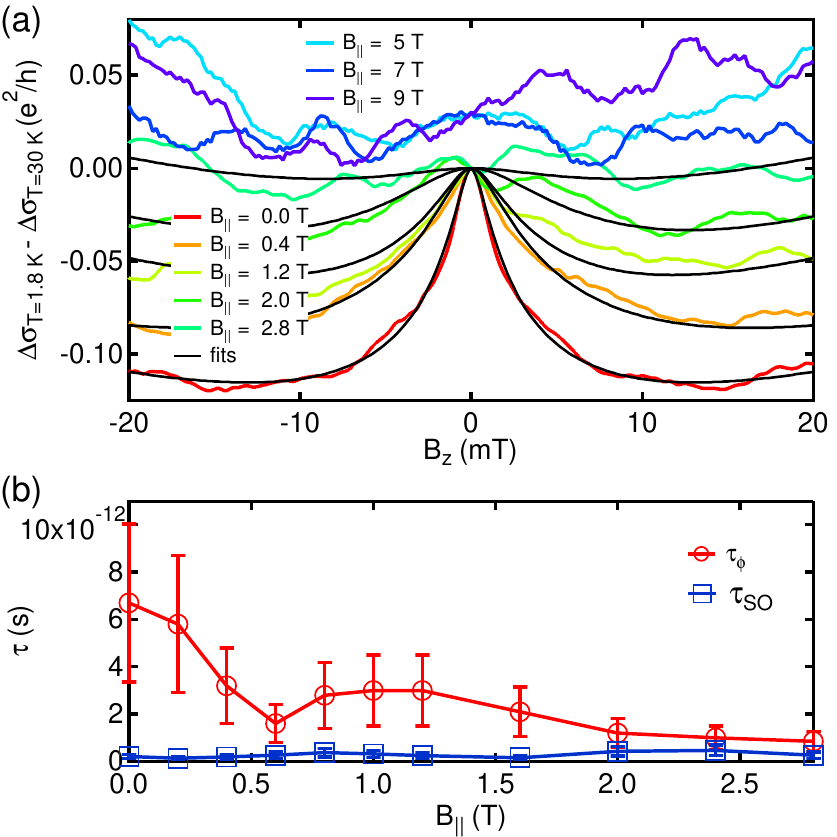}
	\caption{\label{fig:B_ip_dep}\textbf{In-plane magnetic field dependence of device A:} The quantum correction to the magneto conductivity at the CNP and at zero perpendicular electric field is shown for different in-plane magnetic field strengths B$_\parallel$ in \textbf{(a)}. Here, n was averaged in the range of \SIrange{-1e11}{1e11}{\per\square\centi\metre\per\square}. The WAL peak gradually decreases in height and broadens as B$_\parallel$ is increased. The traces at $B_\parallel=$~5, 7, \SI{9}{\tesla} are offset by \SI{0.03}{e^2/h} for clarity. In \textbf{(b)} the extracted phase coherence time $\tau_\phi$ and the total spin-orbit scattering time $\tau_\mathrm{SO}$ are plotted versus B$_\parallel$. $\tau_\phi$ clearly reduces, whereas $\tau_\mathrm{SO}$ remains roughly constant over the full B$_\parallel$ range investigated.}
\end{figure}

The B$_\parallel$ dependence of the quantum correction to the magneto conductivity of device A at the CNP and at zero perpendicular electric field was investigated, as shown in Fig. \ref{fig:B_ip_dep}. The WAL peak decreases and broadens with increasing B$_\parallel$ until it completely vanishes at  B$_\parallel\simeq$\SI{3}{\tesla}. Neither a reappearance of the WAL peak, nor a transition to WL, is observed at higher B$_\parallel$ fields (up to \SI{9}{\tesla}). A qualitatively similar behaviour was observed for device D. Fits with equation \ref{eq:WAL} allow the extraction of $\tau_\phi$ and $\tau_{SO}$, which are shown in Fig.~\ref{fig:B_ip_dep} \textbf{(b)} for B$_\parallel$ fields lower than \SI{3}{\tesla}. A clear decrease of $\tau_\phi$ is observed while $\tau_{SO}$ remains constant.

The reduction in  $\tau_\phi$ with increasing B$_\parallel$ was previously attributed to enhanced dephasing due to a random vector potential created by a corrugated graphene layer in an in-plane magnetic field \cite{2010_Lundeberg}. The clear reduction in $\tau_\phi$ with constant $\tau_{SO}$ and the absence of any appearance of WL at larger B$_\parallel$ also strongly suggests that a similar mechanism is at play here. Therefore, the vanishing WAL peak is due to the loss of phase coherence and not due to the fact that the Zeeman energy ($E_z$) is exceeding the SOC strength. Using the range where WAL is still present, we can define a lower bound of the crossover field when $\tau_\phi$ drops below \SI{80}{\percent} of its initial value, which corresponds to \SI{2}{\tesla} here. This leads to a lower bound of the SOC strength $\lambda_{SOC}\geq E_z\sim$~\SI{0.2}{\milli\eV} given a g-factor of 2.

\section{Discussion}

The effect of SOC was investigated in high quality vdW-heterostructures of WSe$_2$/Gr/hBN at the CNP, as there the effects of SOC are expected to be most important. The two-terminal conductance measurements are not influenced by contact resistances nor pn-interfaces close to the CNP. At larger doping, the two-terminal conductance would need to be considered with care.

Phase coherence times around \SIrange{4}{7}{\pico\second} were consistently found from fits to Eq. \ref{eq:WAL} and from the autocorrelation of UCF. It is commonly known that the phase coherence time is shorter at the CNP than at larger doping \cite{2008_Tikhonenko, 2010_Lundeberg}. Moreover, large diffusion coefficients lead to long phase coherence lengths being on the order of the device size ($l_\phi = \sqrt{D\tau_\phi}\approx$~\SI{1}{\micro\meter}), which in turn leads to large UCF amplitudes making the analysis harder. 

In general Eq.~\ref{eq:WAL} is only applicable for short $\tau_{iv}$. Since $\tau_{iv}$ is unknown in these devices, only an estimate can be given here. WL measurements of graphene on hBN found $\tau_{iv}$ on the order of picoseconds \cite{own_WL_measurement, 2014_Couto}. Inter-valley scattering is only possible at sharp scattering centres as it requires a large momentum change. It is a reasonable assumption that the defect density in WSe$_2$, which is around \SI{1e12}{\per\square\centi\metre} \cite{2016_Addou}, is larger than in the high quality hBN \cite{2007_Taniguchi}. This leads to shorter $\tau_{iv}$ times in graphene placed on top of WSe$_2$ and makes Eq.~\ref{eq:WAL} applicable despite the short spin-orbit scattering times found here. In the case of weaker SOC, Eq.~\ref{eq:WAL} cannot be used. Instead, a more complex analysis including $\tau_{iv}$ and $\tau_*$ is needed. This was used for device D, and is presented in the Supplemental Material. 

Spin-orbit scattering rates were successfully extracted at the CNP and $\tau_{asy}$ was found to be around \SIrange{4}{7}{\pico\second} whereas $\tau_{sym}$ was found to be much shorter, around \SIrange{0.1}{0.3}{\pico\second}. In these systems, if $\tau_{iv}$ is sufficiently short, $\tau_{asy}/2$ is predicted to represent the out-of-plane spin relaxation time $\tau_\perp$ and $\tau_{sym}$ then represents the in-plane spin relaxation time $\tau_\parallel$ \cite{2017_Cummings}. For the time scales stated above, a spin relaxation anisotropy $\tau_\perp/\tau_\parallel \sim 20$ is found (see Supplemental Material for detailed calculation). This large anisotropy in spin relaxation is unique for systems with a strong valley-Zeeman SOC. Similar anisotropies have been found recently in spin valves in similar systems \cite{2017_Ghiasi, 2017_Benitez}.

In order to link spin-orbit scattering time scales to SOC strengths, spin relaxation mechanisms have to be considered. The simple definition of $\hbar/\tau_{SO}$ as the SOC strength is only valid in the limit where the precession frequency is much larger than the momentum relaxation rate (e.g. full spin precession occurs between scattering events). In the following we concentrate on the parameters from device A that were extracted close to the CNP. The dependence on $\tau_p$ in device A can most likely be assumed to be very similar to that observed in device C. Within the investigated density range of \SIrange{-2.5e11}{+2.5e11}{\per\square\centi\metre}, including residual doping, an average Fermi energy of \SI{45}{\milli\eV} was estimated. This is based on the density of states of pristine graphene, which should be an adequate assumption for a Fermi energy larger than any SOC strengths.

The symmetric spin-orbit scattering time $\tau_{sym}$ contains contributions from the intrinsic SOC and from the valley-Zeeman SOC. Up to now, only the intrinsic SOC has been considered in the analysis of WAL measurements, and the impact of valley-Zeeman SOC has been ignored. However, as we now explain, it is highly unlikely that intrinsic SOC is responsible for the small values of $\tau_{sym}$. The intrinsic SOC is expected to relax spin via the Elliott-Yafet (EY) mechanism \cite{2012_Ochoa}, which is given as
\begin{equation}
	\label{eq:EY}
	\tau_s = \left(\frac{2E_F}{\lambda_I}\right)^2 \tau_p,
\end{equation}
where $\tau_s$ is the spin relaxation time, $E_F$ is the Fermi energy, $\lambda_I$ is the intrinsic SOC strength and $\tau_p$ is the momentum relaxation time \cite{2012_Ochoa}. Since the intrinsic SOC does not lead to spin-split bands and hence no spin-orbit fields exist that could lead to spin precession, a relaxation via the Dyakonov-Perel mechanism can be excluded. Therefore, we can estimate $\lambda_I = 2E_F/\sqrt{\tau_{sym}\tau_p^{-1}}\sim$~\SI{110}{\milli\eV} using $\tau_{sym}\sim$~\SI{0.2}{\pico\second}, a mean Fermi energy of \SI{45}{\milli\eV} and a momentum relaxation time of \SI{0.3}{\pico\second}. The extracted value for $\lambda_I$ would correspond to the opening of a topological gap of \SI{220}{\milli\eV}. In the presence of a small residual doping (here \SI{30}{\milli\eV}), such a large topological gap should easily be detectable in transport. However, none of our transport measurements confirm this. In addition, the increase of $\tau_{sym}^{-1}$ with $\tau_p$, as shown in Fig.~\ref{fig:n_dep}, does not support the EY mechanism.

On the other hand, Cummings et al. have shown that the in-plane spins are also relaxed by the valley-Zeeman term via a Dyakonov-Perel mechanism where $\tau_{iv}$ takes the role of the momentum relaxation time \cite{2017_Cummings}:
\begin{equation}
	\label{eq_VZ}
	\tau_s^{-1} = \left(\frac{2\lambda_{VZ}}{\hbar}\right)^2\tau_{iv}.
\end{equation}
While this equation applies in the motional narrowing regime of spin relaxation, our measurement appears to be near the transition where that regime no longer applies. Taking this into consideration (see Supplemental Material), we estimate $\lambda_{VZ}$ to be in the range of \SIrange{0.23}{2.3}{\milli\eV} for a $\tau_{sym}$ of \SI{0.2}{\pico\second} and a $\tau_{iv}$ of \SIrange{0.1}{1}{\pico\second}. This agrees well with first principles calculations \cite{2016_Gmitra}. The large range in $\lambda_{VZ}$ comes from the fact that $\tau_{iv}$ is not exactly known.

Obviously, $\tau_{sym}$ could still contain parts that are related to the intrinsic SOC ($\tau_{sym}^{-1} =\tau_{sym, I}^{-1} + \tau_{sym, VZ}^{-1}$ ). As an upper bound of $\lambda_I$, we can give a scale of \SI{15}{\milli\eV}, which corresponds to half the energy scale due to the residual doping in the system. This would lead to $\tau_{sym, I}\sim$~\SI{10}{\pico\second}. Such a slow relaxation rate ($\tau_{sym, I}^{-1}$) is completely masked by the much larger relaxation rate $\tau_{sym, VZ}^{-1}$ coming from the valley-Zeeman term. Therefore, the presence of the valley-Zeeman term makes it very hard to give a reasonable estimate of the intrinsic SOC strength.

The asymmetric spin-orbit scattering time $\tau_{asy}$ contains contributions from the Rashba-SOC and from the PIA SOC. Since the PIA SOC scales linearly with the momentum, it can be neglected at the CNP. Here, $\tau_{asy}$ represents only the spin-orbit scattering time coming from Rashba SOC. It is known that Rashba SOC can relax the spins via the Elliott-Yafet mechanism \cite{2012_Ochoa}. In addition, the Rashba SOC leads to a spin splitting of the bands and therefore to a spin-orbit field. This opens a second relaxation channel via the Dyakonov-Perel mechanism \cite{1972_Dyakonov}. In principle the dependence on the momentum scattering time $\tau_p$ allows one to distinguish between these two mechanisms. Here, $\tau_{asy}^{-1}$ does not monotonically depend on $\tau_p$ as one can see in Fig.~\ref{fig:n_dep} and therefore we cannot unambiguously decide between the two mechanisms. 

Assuming that only the EY mechanism is responsible for spin relaxation, then $\lambda_R = E_F/\sqrt{4\tau_{asy}\tau_p^{-1}}\sim$~\SI{5.0}{\milli\eV} can be estimated, using $\tau_{asy}$ of \SI{6}{\pico\second}, a mean Fermi energy of \SI{45}{\milli\eV} and a momentum relaxation time of \SI{0.3}{\pico\second}. On the other hand, pure DP-mediated spin relaxation leads to $\lambda_R = \hbar/\sqrt{2\tau_{asy}\tau_p}\sim$~\SI{0.35}{\milli\eV}. The Rashba SOC strength estimated by the EY relaxation mechanism is large compared to first principles calculations \cite{2016_Gmitra}, which agree much better with the SOC strength estimated by the DP mechanism. This is also in agreement with previous findings \cite{2016_Yang, 2017_Yang}.

Since there is a finite valley-Zeeman SOC, which is a result of different intrinsic SOC on the A sublattice and B sublattice, a staggered sublattice potential can also be expected. The presence of a staggered potential, meaning that the on-site energy of the A atom is different from the B atom on average, leads to the opening of a trivial gap of $\Delta$ at the CNP. Since there is no evidence of an orbital gap, we take the first principles calculations as an estimate of $\Delta=$~\SI{0.54}{\milli\eV}.

\begin{figure}[htbp]
	\includegraphics[scale=1]{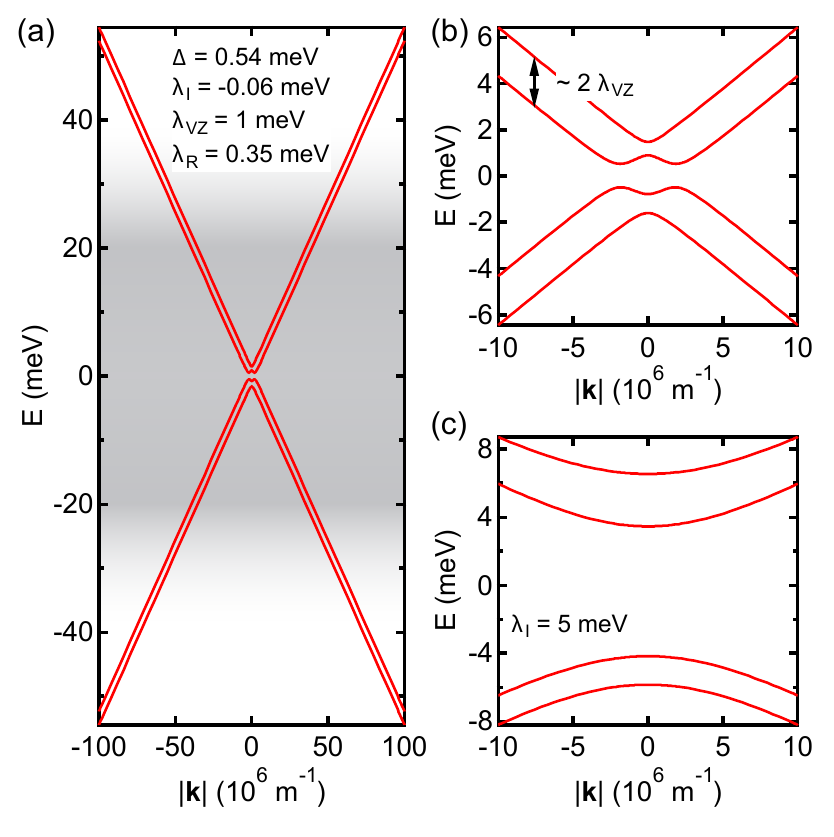}
	\caption{\label{fig:BS}\textbf{Possible low energy band structures:} \textbf{(a)} and \textbf{(b)} show the band structures using the Hamiltonian of Eq.~\ref{eq:Hamiltonian} with the parameters listed in \textbf{(a)}. The unknown parameters $\Delta$ and $\lambda_I$ were taken from Ref. \citep{2016_Gmitra}. In \textbf{(a)}, the band structure is shown in the density range of \SIrange{-2.5e11}{2.5e11}{\per\square\centi\metre} (CNP), which corresponds the the one investigated above. The energy range dominated by charge puddles is indicated by the grey shaded region. \textbf{(b)} shows a zoom in at low energy. In \textbf{(c)}, $\lambda_I$ of \SI{5}{\milli\eV} is assumed to show the changes due to the unknown $\lambda_I$ at low energy.}
\end{figure}

Knowing all relevant parameters in Eq.~\ref{eq:Hamiltonian}, a band structure can be calculated, which is shown in Fig.~\ref{fig:BS}. The bands are spin split mainly due to the presence of strong valley-Zeeman SOC but also due to the weaker Rashba SOC. At very low energies, an inverted band is formed due to the interplay of the valley-Zeeman and Rashba SOC, see Fig.~\ref{fig:BS}~(b). This system was predicted to host helical edge states for zigzag graphene nanoribbons, demonstrating the quantum spin Hall effect \cite{2016_Gmitra}. In the case of stronger intrinsic SOC, which we cannot estimate accurately, a band structure as in Fig.~\ref{fig:BS}~\textbf{(c)} is expected with a topological gap appearing at low energies. We would like to note here, that this system might host a quantum spin Hall phase. However, its detection is still masked by device quality as the minimal Fermi energy is much larger than the topological gap, see also Fig.~\ref{fig:BS}~(a).

Our findings are in good agreement with the calculations by Gmitra et al. \cite{2016_Gmitra}. However, we have to remark that whereas the calculations were performed for single-layer TMDCs, we have used multilayer WSe$_2$ as a substrate. Single-layer TMDCs are direct band-gap semiconductors with the band gap located at the K-point whereas multilayer TMDCs have an  indirect band gap. Since the SOC results from the mixing of the graphene orbitals with the WSe$_2$ orbitals, the strength of the induced SOC depends on the relative band alignment between the graphene and WSe$_2$ band, which will be different for single- or multilayer TMDCs. This difference was recently shown by Wakamura et al. \cite{2017_Wakamura}. Therefore using single-layer WSe$_2$ to induce SOC might even enhance the coupling found by our studies. Furthermore, the parameters taken from Ref.~\cite{2016_Gmitra} for the orbital gap and for the intrinsic SOC therefore have to be taken with care.

\section{Conclusion}
\label{sec:Conclusion}
In conclusion we measured weak anti-localization in high quality WSe$_2$/Gr/hBN vdW-heterostructures at the charge neutrality point. The presence of a clear WAL peak reveals a strong SOC with a much faster spin relaxation of in-plane spins compared to out-of-plane spins. Whereas previous studies have also found a clear WAL signal, we present for the first time a complete interpretation of all involved SOC terms considering their relaxation mechanisms. This includes the finding of a very large spin relaxation anisotropy that is governed by the presence of a valley-Zeeman SOC that couples spin to valley. The relaxation mechanism at play here is very special since it relies on intervalley scattering and can only occur in materials where a valley degree of freedom is present and coupled to spin. This is in excellent good agreement with recent spin-valve measurements that found also very large spin relaxation anisotropies in similar systems \cite{2017_Ghiasi, 2017_Benitez}.

In addition, we investigated the influence of an in-plane magnetic field on the WAL signature. Due to the loss of phase coherence, a lower bound of all SOC strengths of \SI{0.2}{\milli\eV} can be given, which is in agreement with the numbers presented above. This approach does not depend on accurate fitting of WAL peaks nor on the interpretation of spin-orbit scattering rates.

The coupling of spin and valley opens new possibilities in exploring spin and valley degrees of freedom in graphene. In the case of bilayer graphene in proximity to WSe$_2$ an enormous gate tunability of the SOC strength is predicted since full layer polarization can be achieved by an external electric field \cite{2017_Gmitra, 2017_Khoo}. This is just one of many possible routes to investigate in the future.

\textbf{Acknowledgments}

The authors gratefully acknowledge fruitful discussions on the interpretation of the experimental data with Martin Gmitra and Vladimir Fal'ko. Clevin Handschin is acknowledged for helpful discussions on the sample fabrication. This work has received funding from the European Union’s Horizon 2020 research and innovation programme under grant agreement No 696656 (Graphene Flagship), the Swiss National Science Foundation, the Swiss Nanoscience Institute, the Swiss NCCR QSIT and ISpinText FlagERA network OTKA PD-121052 and OTKA FK-123894. P.M. acknowledges support as a Bolyai Fellow. ICN2 is supported by the Severo Ochoa program from Spanish MINECO (Grant No.\ SEV-2013-0295) and funded by the CERCA Programme / Generalitat de Catalunya.

\textbf{Author contributions}
S.Z. fabricated and measured the devices with the help of P.M. K.M. contributed to the fabrication of device C. S.Z. analysed the data with help from P.M. and inputs from C.S.. S.Z., P.M., A.W.C., J.H.G and C.S. were involved in the interpretation of the results. S.Z. wrote the manuscript with inputs from P.M., C.S., A.W.C. and J.H.G., K.W. and T.T. provided the hBN crystals used in the devices.


\end{document}


\title{Supplemental Material: Large spin relaxation anisotropy and valley-Zeeman spin-orbit coupling in WSe2/Gr/hBN heterostructures}


\author{Simon Zihlmann}
\email{Simon.Zihlmann@unibas.ch}
\affiliation{Department of Physics, University of Basel, Klingelbergstrasse 82, CH-4056 Basel, Switzerland}

\author{Aron W. Cummings}
\affiliation{Catalan Institute of Nanoscience and Nanotechnology (ICN2), CSIC and BIST, Campus UAB, Bellaterra, 08193 Barcelona, Spain\\}

\author{Jose H. Garcia}
\affiliation{Catalan Institute of Nanoscience and Nanotechnology (ICN2), CSIC and BIST, Campus UAB, Bellaterra, 08193 Barcelona, Spain\\}

\author{M\'at\'e Kedves}
\affiliation{Department of Physics, Budapest University of Technology and Economics and Nanoelectronics 'Momentum' Research Group of the Hungarian Academy of Sciences, Budafoki ut 8, 1111 Budapest, Hungary}

\author{Kenji Watanabe}
\affiliation{National Institute for Material Science, 1-1 Namiki, Tsukuba, 305-0044, Japan\\}

\author{Takashi Taniguchi}
\affiliation{National Institute for Material Science, 1-1 Namiki, Tsukuba, 305-0044, Japan\\}

\author{Christian Sch\"onenberger}
\affiliation{Department of Physics, University of Basel, Klingelbergstrasse 82, CH-4056 Basel, Switzerland}

\author{P\'eter Makk}
\email{peter.makk@mail.bme.hu}
\affiliation{Department of Physics, University of Basel, Klingelbergstrasse 82, CH-4056 Basel, Switzerland}
\affiliation{Department of Physics, Budapest University of Technology and Economics and Nanoelectronics 'Momentum' Research Group of the Hungarian Academy of Sciences, Budafoki ut 8, 1111 Budapest, Hungary}

\date{\today}
%

\maketitle


\section{Fabrication and measurement details}
\label{sec:FabricationDetails}
Graphene (obtained from natural graphite, NGS), WSe$_2$ (obtained from hQgraphene) and hBN (grown by Taniguchi and Watanabe) were exfoliated with Nitto tape onto Si wafers with \SI{300}{\nano\metre} of SiO$_2$. The WSe$_2$/Gr/hBN vdW-heterostructures were assembled using a dry pick-up method developed by Zomer et al. \cite{2014_Zomer}. After the assembly of the vdW-heterostructures, the stacks were annealed in H$_2$/N$_2$ mixture for \SI{100}{\minute} at \SI{200}{\celsius} to remove polymer residues and to make the stack more homogeneous (merging of bubbles). Higher temperatures were avoided in order not to damage the WSe$_2$ layer. The stacks were shaped into a Hall-bar mesa by standard e-beam lithography and reactive ion etching using a SF$_6$, O$_2$ and Ar-based plasma. One-dimensional side contacts were then fabricated with e-beam lithography and the evaporation of \SI{10}{\nano\metre} Cr and \SI{50}{\nano\metre} Au. Lift-off was performed in warm acetone. In order to insulate the top gate from the exposed graphene at the edge of the mesa, an insulating MgO layer was evaporated before the Ti/Au of the top gate.

Standard low frequency lock-in techniques were used to measure differential conductance and resistance in two- and four-terminal configuration. The samples were measured in a $^3$He system at temperatures down to \SI{0.25}{\kelvin} and in a variable temperature insert at temperature of \SI{1.8}{\kelvin} and higher. The magnetic in-plane field was applied using a vector magnet. The small misalignment of the sample plane with the in-plane magnetic field was compensated by a finite offset field in the out-of-plane direction. This offset was found to scale linearly with the applied in-plane field. 

The back and top gate lever arms ($\alpha_{BG}, \alpha_{TG}$) were found from Hall measurements and the charge carrier density in the graphene was calculated using a simple capacitance model,
\begin{equation}
	\label{eq:density}
	n = \alpha_{BG}\left(V_{BG} - V_{BG}^0\right) + \alpha_{TG}\left(V_{TG} - V_{TG}^0\right),
\end{equation}
where $V_{BG}^0$ and $V_{BG}^0$ account for some offset doping of the graphene. Similarly, the applied electric field (field direction out of plane) was obtained:
\begin{equation}
	\label{eq_E-field}
	E = \frac{1}{d_{BG}}\left(V_{BG} - V_{BG}^0\right) - \frac{1}{d_{TG}}\left(V_{TG} - V_{TG}^0\right),
\end{equation}
where $d_{BG}$ and $d_{TG}$ denote the thickness of the back and top gate dielectric. The thicknesses of the bottom WSe$_2$ flake and the top hBN flake were determined by atomic force microscopy. To account for the residual doping, the density was corrected in the following way: $n_{corr} = \sqrt{n^2 + n_*^2}$. It was the corrected density $n_{corr}$ that was used for the calculation of the diffusion constant via the Einstein relation and for the estimation of the Fermi energy.

\section{Fitting of magneto conductivity data from device B}
\label{sec:FittingDeviceB}
\begin{figure}[htbp]
	\includegraphics[scale=1]{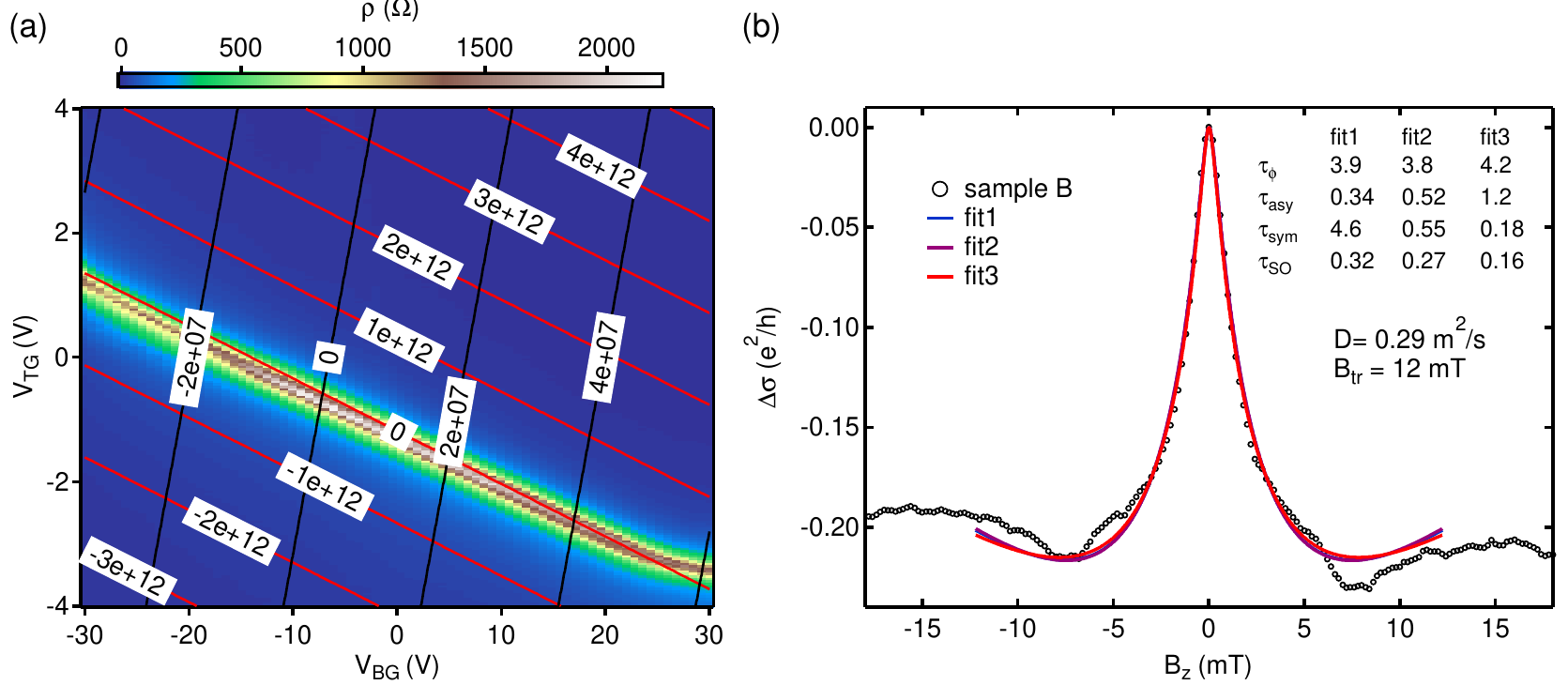}
	\caption{\label{fig:fitting}\textbf{Data from device B:} (a) shows the resistivity as a function of $v_{TG}$ and $V_{BG}$. Constant density contours are indicated with red solid lines and constant electric field contours is solid black lines. (b) shows the quantum correction to the magneto conductivity of device B at zero electric field within a density range of \SIrange{-5e11}{5e11}{\per\square\centi\metre}. The same procedure as described in the main text was used. The results for three different limits are shown and their parameters are indicated. The fitting was restricted to the range of the transport field B$_\mathrm{tr}$ = \SI{12}{\milli\tesla}.}
\end{figure}

As mentioned in the main text, a second device B was investigated as well. A gate-gate map of the resistivity of device B is shown in Fig.~\ref{fig:fitting}~(a). A field effect mobility of \SI{\sim25000}{\square\centi\metre\per\volt\per\second} and a residual doping of \SI{\sim7e10}{\per\square\centi\metre} were found. The quantum correction to the magneto conductivity was measured at the charge neutrality point for different electric fields. The same analysis was performed as mentioned in the main text. The extracted quantum correction to the magneto conductivity was also fit using Eq. 1 from the main text considering the three different cases as elaborated in the main text. Since the quality of device B is higher than that of device A, the diffusion constant is larger and hence the mean free path $l_{mfp}$ is longer. This leads to a much smaller transport field as this scales with $l_{mfp}^{-2}$. Therefore, the fitting range here was limited to \SI{12}{\milli\tesla}, which poses serious limits on the quality of the fit. It is very difficult to independently extract the different spin-orbit scattering times as obviously seen in Fig.~\ref{fig:fitting}, where basically all three fits overlap. Only at larger fields would the three fits be distinguishable. However, the time scales extracted here do not contradict the results presented in the main text. The strength of the total SOC, captured in $\tau_{SO}$, is roughly the same for all three fits. As can be seen in Fig.~\ref{fig:fitting}, the total spin-orbit scattering time $\tau_{SO}$ is more robust with respect to different fitting limits. Therefore, we only consider $\tau_{SO}$ for device B in the next section.

\section{Electric field dependence of the spin-orbit scattering rates}
\label{sec:Efield_dependence}
\begin{figure}[htbp]
	\includegraphics[scale=1]{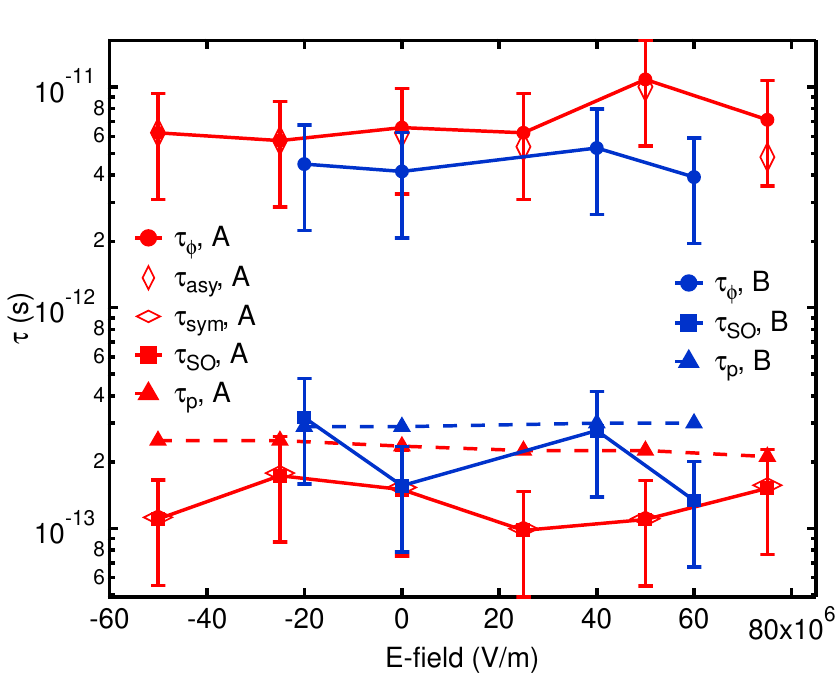}
	\caption{\label{fig:E_field}\textbf{Electric field dependence of device A and B:} The extracted spin-orbit scattering time scales $\tau_{asy}, \tau_{sym}, \tau_{SO}$ and $\tau_\phi$ were extracted for different perpendicular electric field around the charge neutrality point. In addition, the momentum scattering time $\tau_p$ extracted from the diffusion constant is also shown. In the case of device B, only the total spin-orbit scattering time $\tau_{SO}$ is given, as a reliable extraction of $\tau_{asy}$ and $\tau_{sym}$ was not possible in this device (see discussion above).}
\end{figure}

The presence of a top and a back gate in our devices allows us to tune the carrier density and the transverse electric field independently in devices A and B. In the case of device A,  the SOC strength was found to be electric field independent at the CNP in the range of \SIrange{-5e7}{8e7}{V/m} as shown in Fig~\ref{fig:E_field}. The electric field range was limited by the fact that at large positive gate voltages the Fermi energy was shifted into the conduction band of the WSe$_2$ whereas at large negative gate voltages gate instabilities occurred. Within the investigated electric field range $\tau_{asy}$ was found to be in the range of \SIrange{5}{10}{\pico\second}, always close to $\tau_\phi$. $\tau_{sym}$ on the other hand was found to be around \SIrange{0.1}{0.3}{\pico\second} while $\tau_p$ was around \SIrange{0.2}{0.3}{\pico\second} for device A. The total spin-orbit scattering time $\tau_{SO}$ is mostly given by $\tau_{sym}$. Device B, where only $\tau_\phi$ and $\tau_{SO}$ could be extracted reliably, shows similar results as device A. Therefore, we conclude that the in the electric field range \SIrange{-5e7}{8e7}{V/m} no tuning of the SOC strength with electric field is observed. From first principles calculations, the Rashba SOC is expected to change by \SI{10}{\percent} if the electric field is tuned by \SI{1e9}{\volt\per\metre} and also the intrinsic and valley-Zeeman SOC parameters are expected to change slightly \cite{2016_Gmitra}. However, within the resolution of the extraction of the spin-orbit scattering time scales, we cannot establish a clear trend.

These findings are in contrast to previous studies that found an electric field tunability of $\tau_{asy}$ and $\tau_{SO}$ on a similar electric field scale in graphene/WSe$_2$ devices \cite{2017_Voelkl}. However, it is important to note that the changes are small and since no error bars are given, it is hard to tell if the three data points show a clear trend. Another study found a linear tunability of $\tau_{asy}$ of roughly \SI{10}{\percent} on a similar electric field scale in graphene/WS$_2$ devices \cite{2016_Yang}. There, $\tau_{sym}$ was neglected with the argument that it cannot lead to spin relaxation.  However, it was shown that $\tau_{sym}$ can lead to spin relaxation \cite{2017_Cummings} and therefore it cannot be neglected in the analysis. In our case, it is the dominating spin relaxation mechanism.

\section{Spin relaxation anisotropy}
\label{sec:anisotropy}
Cummings et al. have found a giant spin relaxation anisotropy in systems with strong valley-Zeeman SOC that is commonly found in graphene/TMDC heterostructures \cite{2017_Cummings}. They derived the following equation:
\begin{equation}
\frac{\tau_\perp}{\tau_\parallel} = \left(\frac{\lambda_{VZ}}{\lambda_R}\right)^2\frac{\tau_{iv}}{\tau_p} + 1/2
\end{equation}
where $\tau_\perp$ is the out-of-plane spin relaxation time, $\tau_\parallel$ the in-plane spin relaxation time, $\lambda_{VZ}$ is the SOC strength of the valley-Zeeman SOC, $\lambda_R$ is the SOC strength of the Rashba SOC and $\tau_{iv}$ and $\tau_p$ represent the intervalley and momentum scattering times respectively. If a strong intervalley scattering is assumed, which is a prerequisite for the application of the WAL theory \cite{2012_McCann}, $\tau_\perp$ is given by $\tau_{asy}/2$ and $\tau_\parallel$ is given by $\tau_{sym}$. We therefore get a spin relaxation anisotropy $\tau_\perp / \tau_\parallel \approx \tau_{asy}/2\tau_{sym} \approx 20$, which is much larger than what is expected for usual 2D Rashba systems. Furthermore, assuming a ratio of $\tau_{iv}/\tau_p \approx 1$, which corresponds to very strong intervalley scattering, a ratio of $\lambda_{VZ}/\lambda_R\approx 6$ is expected.

\section{Estimate of valley-Zeeman SOC strength}
For a valley-Zeeman SOC strength $\lambda_{VZ}$, the spin splitting is $2\lambda_{VZ}$ and the precession frequency is $\omega = 2\lambda_{VZ}/\hbar$. In the D'yakonov-Perel' (DP) regime of spin relaxation, when $\omega \tau_{iv} < 1$, the in-plane spin relaxation rate is $\tau_{s\parallel}^{-1} = (2\lambda_{VZ}/\hbar)^2 \tau_{iv}$. However, if $\omega \tau_{iv} > 1$, then the spin can fully precess before scattering randomizes the spin-orbit field, and the spin lifetime scales with the intervalley time, $\tau_{s\parallel} = 2\tau_{iv}$. A plot of these two regimes is shown below, where we have taken our derived limits of $\lambda_{VZ} = 0.23$ and \SI{2.3}{\milli\eV} (see below) as well as the DFT-derived value of \SI{1.19}{\milli\eV}.

\begin{figure}[h]
\includegraphics{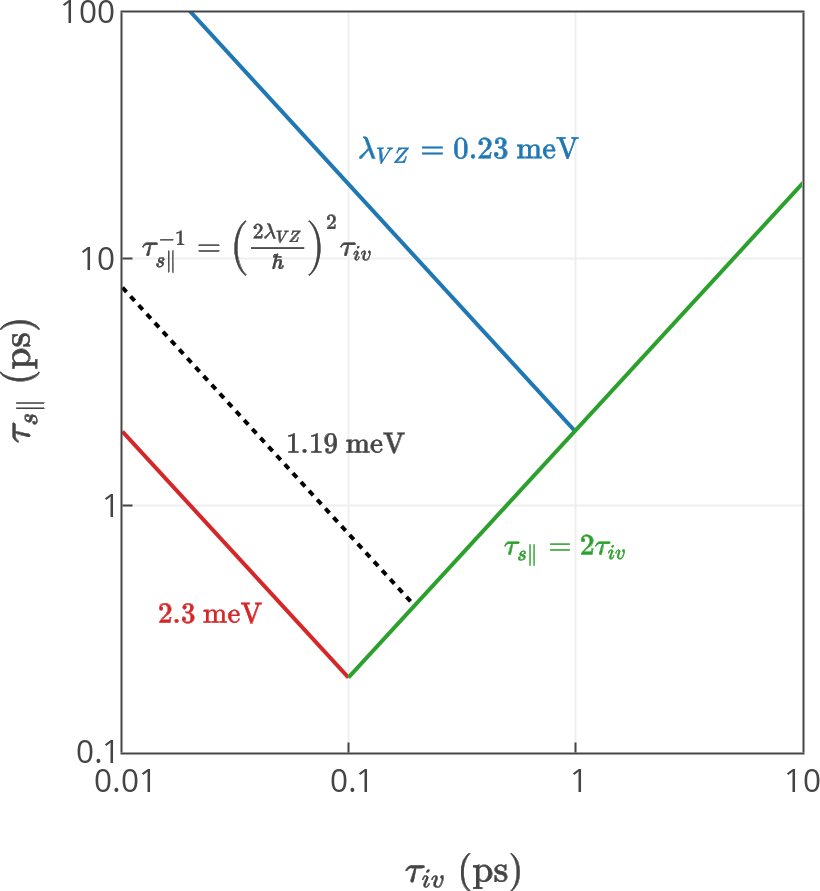}
\caption{Dependence of in-plane spin relaxation time $\tau_{s\parallel}$ on intervalley scattering time $\tau_{iv}$. Red and blue lines show the dependence in the DP regime of spin relaxation, for the largest and smallest estimated values of $\lambda_{VZ}$. The black dashed line show the value derived from DFT \cite{2016_Gmitra}. The green line shows the dependence in the coherent spin precession regime.}
\end{figure} \label{fig_valleyzeeman}

Considering this behavior, the condition $\tau_{s\parallel} \ge 2\tau_{iv}$ should always be satisfied. Meanwhile, our measurements revealed $\tau_{sym} = 0.2$ ps and $\tau_{iv} \approx 0.1-1$ ps, which violates this condition for all except the smallest value of $\tau_{iv}$. One way to account for this is to consider the impact of spin-orbit disorder on the in-plane spin lifetime. Assuming that the $\tau_{s\parallel}$ from uniform valley-Zeeman SOC is given by $2\tau_{iv}$, and the rest comes from spin-orbit disorder, we can estimate an upper bound of $\lambda_{VZ} = \hbar / \sqrt{4 (2\tau_{iv}) \tau_{iv}} =$~\SIrange{0.23}{2.3}{\milli\eV}.

Another possibility is that since our measurements are right around the transition point $\omega \tau_{iv} = 1$, we could be extracting the in-plane spin precession frequency; $\tau_{sym}^{-1} = \omega$. Doing so would give $\lambda_{VZ} = \hbar / 2\tau_{sym} =$~\SI{1.6}{\milli\eV}, which fits in the range derived above. Overall, since the experiments appear to be close to this transition point, all methods of deriving the strength of $\lambda_{VZ}$ tend to give similar values, from a few tenths up to a few meV depending on the estimate of $\tau_{iv}$. We would like to note that it is not fully understood how the spin precession frequency enters into the WAL correction and how the corresponding SOC strength would then be extracted. Therefore, further theoretical work is needed.

\section{Data from device D}
\label{sec:dev_D}
\begin{figure}[htbp]
	\includegraphics[scale=1]{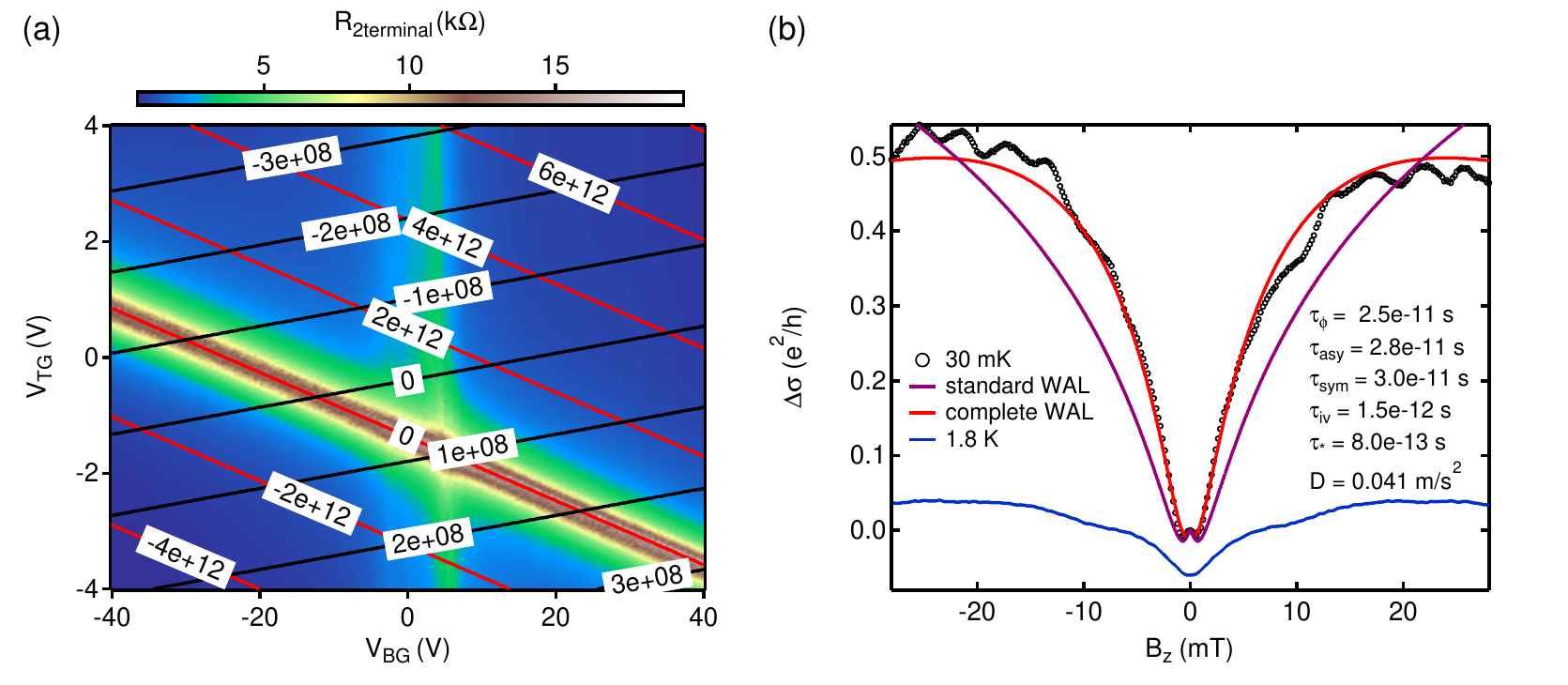}
	\caption{\label{fig:device_D}\textbf{Data from device D:} (a) shows a gate-gate map of the two-terminal resistance of device D. Constant density (red solid line, in units of \si{\per\square\centi\metre}) and electric field (black solid lines, in units of \si{\volt\per\metre}) lines are superimposed on top of that. (b) shows the quantum quantum correction of the magneto conductivity at zero electric field in the density range of \SIrange{-5e11}{5e11}{\per\square\centi\metre}. It shows a WL dip with a tiny feature of WAL around zero $B_z$ at a temperature of \SI{30}{\milli\kelvin}. A possible fit (red) and its parameters, including the influence of $\tau_{iv}$ and $\tau_*$, are indicated. The low magnetic field range can be reasonably well described by the standard WAL formula without $\tau_{iv}$ and $\tau_*$. As a comparison, the magneto conductivity is also shown at \SI{4}{\kelvin}. This trace is vertically offset by \SI{-0.06}{\square e/h} for clarity.}
\end{figure}

The third sample with device D is a WSe$_2$/Gr/hBN stack with a very thin WSe$_2$ (\SI{3}{\nano\metre}) as substrate. The gate-gate map of the two terminal resistance is shown in Fig.~\ref{fig:device_D}~(a). Due to the very thin bottom WSe$_2$ the mobility in this device is around \SI{50000}{\square\centi\metre\per\volt\per\second} and a residual doping of \SI{2.5e11}{\per\square\centi\metre} is found. A typical magneto conductivity trace of this device is shown in Fig.~\ref{fig:device_D}. Mostly, positive magneto conductivity is observed with only a very small feature that shows negative magneto conductivity at \SI{30}{\milli\kelvin}, which was absent at \SI{1.8}{\kelvin}. The magneto conductivity of device D could not be fitted with the standard WAL formula presented in the main text. However, similar curve shapes could be reproduced by including the influence of $\tau_{iv}$ and $\tau_*$. A complete formula can be derived from equation 9 of Ref.~\cite{2012_McCann}. If all relaxation gaps are included and if disorder SOC is neglected one arrives at the following form:
\begin{equation}
	\label{eq:WAL_complete}
	\begin{split}
	\Delta\sigma (B) = -\frac{e^2}{2\pi h}\left[ F \left( \frac{\tau_B^{-1}}{\tau_\phi^{-1}} \right) - F \left( \frac{\tau_B^{-1}}{\tau_\phi^{-1} + 2\tau_{asy}^{-1}} \right) - 2F \left( \frac{\tau_B^{-1}}{\tau_\phi^{-1} + \tau_{asy}^{-1} + \tau_{sym}^{-1}} \right)\right. \\
	\left. - F \left( \frac{\tau_B^{-1}}{\tau_\phi^{-1} + 2\tau_{iv}^{-1}} \right) - 2F \left( \frac{\tau_B^{-1}}{\tau_\phi^{-1} + \tau_{*}^{-1}} \right)\right. \\
	\left. + F \left( \frac{\tau_B^{-1}}{\tau_\phi^{-1} + 2\tau_{iv}^{-1} + 2\tau_{asy}^{-1}} \right) + 2F \left( \frac{\tau_B^{-1}}{\tau_\phi^{-1} + \tau_{*}^{-1} + 2\tau_{asy}^{-1}} \right)\right. \\
	\left.+ 2F \left( \frac{\tau_B^{-1}}{\tau_\phi^{-1} + 2\tau_{iv}^{-1} + \tau_{asy}^{-1} + \tau_{sym}^{-1}} \right) + 4F \left( \frac{\tau_B^{-1}}{\tau_\phi^{-1} + \tau_{*}^{-1} + \tau_{asy}^{-1} + \tau_{sym}^{-1}} \right)\right].
	\end{split}
\end{equation}
However, the addition of two more parameters makes it very hard to unambiguously extract all parameters exactly. Therefore, we do not extract any spin-orbit time scales from this device. The influence of $\tau_{iv}$ and $\tau_*$ are much weaker for the data presented in the main text.

The long phase coherence time $\tau_\phi$~\SI{\sim25}{\pico\second} is attributed to the lower temperature (T=~\SI{30}{\milli\kelvin}) at which the measurement was performed. At higher temperature (\SI{1.8}{\kelvin}), the phase coherence is significantly shorter \SI{\sim4} {\pico\second}(broader dip and reduced overall correction) and the influence of the SOC on the magneto conductivity (WAL) is not observed any longer. 

Both $\tau_{asy}$ and $\tau_{sym}$ seem to be very close to $\tau_\phi$ in sample D. In particular, $\tau_{sym}$ is much longer than in the devices presented in the main text. We conclude that even though there is some indication of SOC in sample D, its overall strength must be smaller than in the devices presented in the main text. Certainly the SOC relevant for $\tau_{sym}$ must be smaller as this time scale is two orders of magnitude longer than in device A and B. This large difference cannot be explained by the shorter $\tau_p$ that is roughly a factor of 5 shorter in device D than in device A and B.

\section{Influence of WSe$_2$ quality}
In addition to WSe$_2$ obtained from hQ graphene, we also investigated devices with WSe$_2$ obtained from Nanosurf as an alternative source. In general, devices with WSe$_2$ from Nanosurf showed more gate instabilities. Some devices showed mobilities around \SI{20000}{\square\centi\metre\per\volt\per\second}. Magneto conductivity was measured in order to investigate possible enhanced SOC, but in none of the devices we did we find a pronounced WAL signature. Some devices showed signatures of WL, whereas some did not show any clear magneto conductivity. For some devices it was impossible to measure magneto conductivity as the devices were not stable enough.
